\documentclass[prd,nopacs,nokeys]{revtex4}
\usepackage{amsfonts}
\usepackage{amsmath}
\usepackage{amssymb}
\usepackage{bbm}
\usepackage[utf8]{inputenc}
\usepackage[caption=false]{subfig}
\usepackage{tensor}
\usepackage{slashed}
\usepackage[centertableaux ]{ytableau}
\usepackage{hyperref}
\usepackage{natbib}
\usepackage{enumerate}
\usepackage{braket,mleftright}
\usepackage{graphicx}
\usepackage{cleveref}
\usepackage{tikz}
\usepackage{tikz,ifthen}

\usetikzlibrary{trees}
\usetikzlibrary{decorations.pathmorphing}
\usetikzlibrary{decorations.markings}
\tikzset{beamerprimary/.style={structure.fg, thick}}
\tikzset{beamersecondary/.style={structure.bg, thick}}
\tikzset{ boson/.style={decorate, decoration={snake}},
     gauge/.style={decorate,decoration={snake,post length=1mm}}  ,
     fermion/.style={postaction={decorate},
        decoration={markings,mark=at position .55 with {\arrow{>}}}},
    fermionloop/.style={postaction={decorate},
        decoration={markings,mark=at position .25 with {\arrow{<}}}}, 
    gluon/.style={decorate, 
        decoration={coil,amplitude=4pt, segment length=5pt}},
    scalar/.style={dashed},
    graviton/.style={double}
,
    dm/.style={double,postaction={decorate},decoration={markings,mark=at position .55 with {\arrow{>}}}}
}

\input{tcilatex}

\begin{document}

\title{Spin portal to dark matter.}
\author{H. Hern\'{a}ndez-Arellano$^{(1)}$, M. Napsuciale$^{(1)}$, S. Rodr\'{\i}guez$^{(2)}$}

\address{$^{1}$ Departamento de F\'{\i}sica, Universidad de
  Guanajuato, Lomas del Bosque 103, Fraccionamiento Lomas del
  Campestre, 37150, Le\'{o}n, Guanajuato, M\'{e}xico,} 
\address{$^{2}$ Facultad de Ciencias F\'isico-Matem\'aticas,
  Universidad Aut\'onoma de Coahuila, Edificio A, Unidad
  Camporredondo, 25000, Saltillo Coahuila M\'exico.} 
  
\begin{abstract}
{\ In this work we study the possibility that dark matter fields transform
in the $(1,0)\oplus(0,1)$ representation of the Homogeneous Lorentz Group.
In an effective theory approach, we study the lowest dimension interacting
terms of dark matter with standard model fields, assuming that dark matter
fields transform as singlets under the standard model gauge group. There are
three dimension-four operators, two of them yielding a Higgs portal to dark
matter. The third operator couple the photon and $Z^0$ fields to the higher
multipoles of dark matter, yielding a \textit{spin portal} to dark matter.
For dark matter ($D$) mass below a half of the $Z^0$ mass, the decays $Z^0\to 
\bar{D} D$ and $H\to \bar{D}D$ are kinematically allowed and contribute to
the invisible widths of the $Z^0$ and $H$. We calculate these decays and use
experimental results on these invisible widths to constrain the values of
the low energy constants finding in general that effects of the spin portal
can be more important that those of the Higgs portal. We calculate the dark
matter relic density in our formalism, use the constraints on the low energy
constants from the $Z^0$ and $H$ invisible widths and compare our results
with the measured relic density, finding
that dark matter with a $(1,0)\oplus(0,1)$ space-time structure must have a 
mass $M>43 ~ GeV$. }
\end{abstract}

\maketitle

\section{Introduction.}


The elucidation of the nature of dark matter is one of the most important
problems in high energy physics \cite{Arcadi:2017kky}. Although dark matter
gravitational effects were noticed during the first half of the last century  \cite{Zwicky:1933gu}
and from recent precise measurements of the cosmic background radiation we know that it 
accounts for around $26\%$ \cite{Ade:2015xua} of the matter-energy content of the universe, 
an identification of dark matter properties is still lacking and a lot of experimental effort is presently
being pursued in order to directly or indirectly detect dark matter
particles, based mainly in the WIMP paradigm \cite{Steigman:1984ac} . The latter is
based on the fact that the proper description of the measured dark matter relic density, 
$\Omega^{exp}_{DM} h^2 =0.1186\pm0.0020 $ \cite{Ade:2015xua,Patrignani:2016xqp},  requires dark
matter to have annihilation cross sections into standard model particles of
the order of those produced by the weak interactions.

From the particle physics side, dark matter is a challenging problem since
there is no particle in the standard model which can be identified with dark
matter and, although some extensions of the standard model such as supersymmetric
models or extra-dimension models have candidates to dark matter, no signal for these 
particles has been found in the
exhaustive search for signals of physics beyond the standard model or direct
search for dark matter signals carried out at the LHC during the past few
years \cite{Varnes:2016rzg,Charlton:2017wfj,Camporesi:2016fjj}.

The problem has also been considered in a model independent way using
effective field theories, where the low energy effects of the unknown theory
at high energies are considered in a systematic expansion, based on general
principles. Effective theories for scalar \cite%
{Silveira:1985rk,McDonald:1993ex,Burgess:2000yq,Kanemura:2010sh,Andreas:2010dz,Djouadi:2011aa,Mambrini:2011ik,Djouadi:2012zc}
, fermion \cite%
{Kanemura:2010sh,Djouadi:2011aa,LopezHonorez:2012kv,Djouadi:2012zc} or
vector \cite{Kearney:2016rng,Bambhaniya:2016cpr,Cotta:2012nj} particles have been
proposed, and several experimental direct searches are motivated by these
formalisms.

The standard model contains spin 1/2 fermions (quarks and leptons), spin 1
bosons (gauge bosons) and a spin 0 boson (the Higgs particle) with the
corresponding fields transforming in the $(\frac{1}{2},0)\oplus (0,\frac{1}{2%
})$, $(\frac{1}{2},\frac{1}{2})$ and $(0,0)$ representations of the
Homogeneous Lorentz Group (HLG) respectively and it is natural that
effective theories so far formulated for dark matter consider dark matter
transforming in these representations.

Recently, the quantum field theory of spin one massive particles
transforming in the $(1,0)\oplus (0,1)$ representation of the HLG
 (spin-one matter fields), was studied in detail in \cite{Napsuciale:2015kua},
 where the field is described by a six-component spinor, similar to the 
 four-component Dirac spinor describing spin $1/2$ fermions. 
It was shown there that a consistent quantum field
theory of spin-one matter fields requires a constrained dynamics formalism
but the constraints are second class and can be solved along Dirac
conventional method \cite{Dirac:1964}. In order to solve the constraints,
however, we need to know the algebraic structure of a covariant basis for
the operators acting in the $(1,0)\oplus (0,1)$ representation space, which
was previously worked out in \cite{Gomez-Avila:2013qaa}. This basis
naturally contains a chirallity operator, $\chi$, and spin-one matter fields
can be decomposed into chiral components transforming in the $(1,0)$ (right)
and $(0,1)$ (left) representations. However, the kinetic term in the free
Lagrangian is not invariant under independent chiral transformations, therefore spin-one
matter fields cannot have linearly realized chiral gauge interactions, hence
they cannot have weak interactions. Nonetheless, it is possible to have
vector-like interactions like $U(1)_Y$ or $SU(3)_c$ standard model
interactions. In addition, spin-one matter fields can have naively
renormalizable self-interactions classified also in \cite{Napsuciale:2015kua}. 

In this work we study the possibility of a $(1,0)\oplus (0,1)$ space-time structure 
for dark matter fields.  Clearly, dark matter with standard model charges
would give sizable contributions to precision measurements of standard model 
observables, thus we assume in this work that dark matter fields transform as 
singlets of the standard model gauge group.

The paper is organized as follows. In the next section we review the
elements of the quantum field theory of spin one matter fields needed for
the calculation of the required cross sections. In Section III we discuss
the leading terms in the effective field theory. In section IV we study the
mass region $M<M_{Z}/2$, calculate the decay width for $Z^0\to \bar{D} D$
and $H\to \bar{D}D$ and find the constraints on the low energy constants
from the $Z^{0}$ and Higgs invisible widths. Section V contains an analysis
of the dark matter relic density in this formalism, when these constraints are
taken into account. Finally, we give our conclusions and perspectives in
section VI and close with an appendix with the required trace calculations 
for operators in the $(1,0)\oplus(0,1)$ representation space.


\section{Quantum field theory for spin-one matter fields: brief review}

In the standard model, matter is described by Dirac fermions which transform
in the $(1/2,0)\oplus (0,1/2)$ representation of the HLG. Spin-one matter
fields are the generalization of Dirac construction to $j=1$, i.e. fields
transforming in the $(1,0)\oplus (0,1)$. The basic object is a six-component
`spinor` $\psi (x)$ and the corresponding quantum field theory was studied
in \cite{Napsuciale:2015kua}, taking advantage of the general construction
of a covariant basis for $(j,0)\oplus(0,j)$ representation space introduced
in \cite{Gomez-Avila:2013qaa}. For $j=1$ the covariant basis is given by the
set of $6\times 6$ matrices $\{ \mathbf{1}, \chi, S^{\mu\nu}, \chi
S^{\mu\nu}, M^{\mu\nu}, C^{\mu\nu\alpha\beta } \}$ where $\chi$ is the
chirality operator, $S^{\mu\nu}$, stands for a symmetric traceless ($S%
\indices{^\mu_\mu}=0$) matrix tensor transforming in the $(1,1)$
representation of the HLG, $M^{\mu\nu}$ are the
HLG generators and $C^{\mu\nu\alpha\beta}$ is a matrix tensor transforming
in the $(2,0)\oplus(0,2)$ representation of the HLG. 

The spin-one matter field is written as 
\begin{equation}
\psi(x)=\sum_{\lambda}\int \frac{d^3 p}{\sqrt{(2\pi)^3 2E}} [ a_{\lambda}(p)%
\mathcal{U}(p,\lambda) e^{-ip.x} + b^{\dagger}_{\lambda}(p)\mathcal{V}%
(p,\lambda)e^{ip.x} ]
\end{equation}
where $\mathcal{U}(p,\lambda)$ ($\mathcal{V}(p,\lambda)$) stands for the
particle (antiparticle) solution with polarization $\lambda$ respectively.
In contrast with the Dirac case, spin-one matter particle and antiparticle
have the same parity. These solutions satisfy 
\begin{equation}
\sum_{\lambda} \mathcal{U}\left( p, \lambda\right) \mathcal{\bar{U}}\left(
p, \lambda \right) = \frac{S\left( p \right) +M^{2}}{2M^{2}} , \qquad
\sum_{\lambda} \mathcal{V}\left( p, \lambda\right) \mathcal{\bar{V}}\left(
p, \lambda \right) = \frac{S\left( p \right) +M^{2}}{2M^{2}} .
\end{equation}
where $S\left( p \right) \equiv S^{\mu\nu} p_{\mu} p_{\nu}$.

The spin-one matter fields free Lagrangian is given by 
\begin{equation}
\mathcal{L}= \frac{1}{2}\partial^\mu \bar{\psi} (x)(g_{\mu\nu}+S_{\mu\nu})
\partial^\nu\psi (x) -m^{2} \bar{\psi} (x)\psi (x).
\end{equation}
where $\bar{\psi} (x)\equiv (\psi (x))^{\dagger}S^{00}$. The $S^{\mu\nu}$
operators satisfy the following anti-commutation relations 
\begin{equation}
\lbrace S^{\mu\nu},S^{\alpha\beta} \rbrace =\frac{4}{3}\left(
g^{\mu\alpha}g^{\nu\beta}+g^{\nu\alpha}g^{\mu\beta}-\frac{1}{2}%
g^{\mu\nu}g^{\alpha\beta}\right) - \frac{1}{6}\left(
C^{\mu\alpha\nu\beta}+C^{\mu\beta\nu\alpha}\right) .
\end{equation}
Further algebraic relations of the operators in the covariant basis and the
connection with the traces needed for the calculations in this work are
deferred to an appendix. The propagator for spin-one matter particles is
given by 
\begin{equation}
i\pi(p)= i \frac{S(p)-p^{2}+2M^2}{2M^2(p^2-M^2+i\varepsilon)}.
\end{equation}

An important outcome of this formalism is that the free field Lagrangian can
be decomposed in terms of the chiral components as 
\begin{equation}
\mathcal{L}= \frac{1}{2} \partial^{\mu}\overline{\psi_{R}}%
\partial_{\mu}\psi_{L}+ \frac{1}{2} \partial^{\mu}\overline{\psi_{R}}
S_{\mu\nu} \partial_{\nu}\psi_{R} -m^{2} \overline{\psi_{R}}\psi_{L} +
R\leftrightarrow L,
\end{equation}
where 
\begin{equation}
\psi_{R}=\frac{1}{2}\left( 1+\chi\right) \psi, \qquad \psi_{L}=\frac{1}{2}%
\left(1-\chi\right)\psi.
\end{equation}
The right (left) field $\psi_{R}$ ($\psi_{L}$) transforms in $(1,0)$ ($(0,1)$) representation of the
HLG. Notice that in the massless case, the kinetic term couples right and
left components, hence it is not invariant under independent chiral
transformations. Therefore, spin-one matter fields cannot have chiral gauge
interactions, although they admit vector gauge interactions. Concerning the
standard model interactions, spin-one matter fields can have only $U(1)_{Y}$
or $SU(3)_{C}$ gauge interactions but not $SU(2)_{L}$ interactions, or
simply be standard model singlets. This result motivate us to explore the
possibility that dark matter be described by spin-one matter fields and we
start with the simplest and most likely possibility: spin-one dark matter fields
transforming as singlets under the standard model gauge group.


\section{Dark matter as spin-one matter fields: effective theory.}


If we consider dark matter as spin-one matter fields (spin-one dark matter
fields in the following) transforming as singlets under the standard model
group, dark matter does not feel the standard model charges. On the other
side, if we have more than one dark matter field, dark matter can have gauge
interactions with its own (vector-like) dark gauge group. In the following we
will assume a simple $U(1)_D$ structure for the dark gauge group, but the
generalization of our results to $SU(N)_D$ is straightforward. We remark
that the only effect of this dark gauge structure in this work is to provide
to dark matter particles with dark charges distinguishing particles from
anti-particles and preventing the direct decay of a dark matter particle
into standard model ones.

At high energies, the standard model and dark sectors couple in a yet
unknown way but the low energy effects of such theory  can be classified in
an expansion in derivatives of the fields. Each term in this expansion has a
low energy constant and the  importance at low energies of each term depends
on the dimension of the corresponding operator, in such a way that the  most
important effects are given by the lowest dimension operators.

The Lagrangian must be a complete scalar operator and if dark matter fields
are standard model singlets (and standard model fields are singlets of the
dark gauge group) the only possibility to have a scalar interacting
Lagrangian is that it be composed of products of singlet operators on both
sides. The construction of the lowest dimension interacting operators in
this case, requires to classify the singlet operators in both sectors. The
most general form of this interaction is 
\begin{equation}
\mathcal{L}_{int}= \sum_{n}\frac{1}{\Lambda^{n-4}}\mathcal{O}_{SM} \mathcal{O%
}_{DM}
\end{equation}
where $\Lambda$ is an energy scale compensating the dimension $n$ of the
product of the standard model singlet operators $\mathcal{O}_{SM}$
constructed with standard model fields and $\mathcal{O}_{DM}$ made of 
spin-one dark matter fields.

It is easy to convince one-self that the lowest dimension standard model
singlet operators are $\tilde{\phi}\phi$ and $B_{\mu\nu}$, where $\phi$
stands for the standard model Higgs doublet and $B_{\mu\nu}$ denotes the 
$U(1)_{Y}$ stress tensor.  Indeed, $\tilde{\phi}\phi$ is simply the singlet
of the $\mathbf{2}\otimes \mathbf{2}$ product of $SU(2)_L$ ( and also a
singlet  under $SU(3)$ and $U(1)_{Y}$), while in general under $SU(N)$ gauge
transformations $U(x)$, the stress (matrix) tensor  operator transforms as 
\begin{equation}
F^{\mu\nu} \to U(x) F^{\mu\nu} U^{-1}(x),
\end{equation}
being strictly invariant only in the $U(1)$ case, thus, in the standard
model, the $U(1)_{Y}$ stress tensor $B_{\mu\nu}$ is a singlet under the  standard 
model gauge group. Singlet operators made of fermion fields or other
combinations can also be constructed but they are higher dimension.

For spin-one matter fields with a dark gauge group $U(1)_D$ , the lowest
dimension operators transforming as standard model and dark gauge group
singlets are of the form $\bar{\psi} O \psi$ where $O$ is one of the $36$
matrix operators in the covariant basis $\{ \mathbf{1}, \chi,
S^{\mu\nu},\chi S^{\mu\nu}, M^{\mu\nu}, C^{\mu\nu\alpha\beta } \}$. These
operators are dimension two and using the symmetry properties of $S^{\mu\nu}$
and $C^{\mu\nu\alpha\beta}$ it is easy to show that the leading interacting
terms in the effective theory are given by 
\begin{equation}
\mathcal{L}_{int}= \bar\psi (g_{s} \mathbf{1}+ i g_{p}\chi) \psi \tilde{\phi}
\phi + g_{t} \bar\psi M_{\mu\nu} \psi B^{\mu\nu},  \label{Leff}
\end{equation}
with low energy constants $g_{s} $, $g_{p} $ and $g_{t} $. There is an
effective Higgs portal to dark matter interactions with standard model
particles given by the first two terms, the second one violating parity. The
third term is an effective interaction coupling dark matter to the photon
and the $Z^0$ boson. Notice however that this interaction does not involve
the weak charges (operators are standard model singlets), but proceeds
through the coupling of the photon and $Z^{0}$ fields to the higher multipoles
(magnetic dipole moment and electric quadrupole moment) of the dark matter,
thus we name it \textit{spin portal} to dark matter. In addition to the
interactions in Eq.(\ref{Leff}) we have the dimension four self-interactions
described in \cite{Napsuciale:2015kua} which are not relevant for the
purposes of this paper.

\begin{figure}[t]
\begin{tikzpicture}
\draw[dm,thick] (-5.5,1) --  (-4,0) ;
\draw[dm,thick] (-5.5,-1) --   (-4,0) ;
\draw[->,gauge,thick] (-4,0) -- node[above]{$k, \mu$} node[below]{$\gamma $}  (-2.5,0) ;
\node at(-0.5,0){$=2g_{t} \cos \theta_{W} M^{\mu\nu} k_{\nu} $};
\draw[dm,thick] (1.5,1) -- (3,0) ;
\draw[dm,thick] (1.5,-1) --  (3,0) ;
\draw[->,gauge,thick] (3,0) -- node[above]{$k, \mu $} node[below]{$Z $} (4.5,0) ;
\node at(6.5,0){$=-2g_{t} \sin \theta_{W} M^{\mu\nu} k_{\nu} $};
\end{tikzpicture}
\caption{Feynman rules from the leading terms in the effective theory. }
\label{FR}
\end{figure}
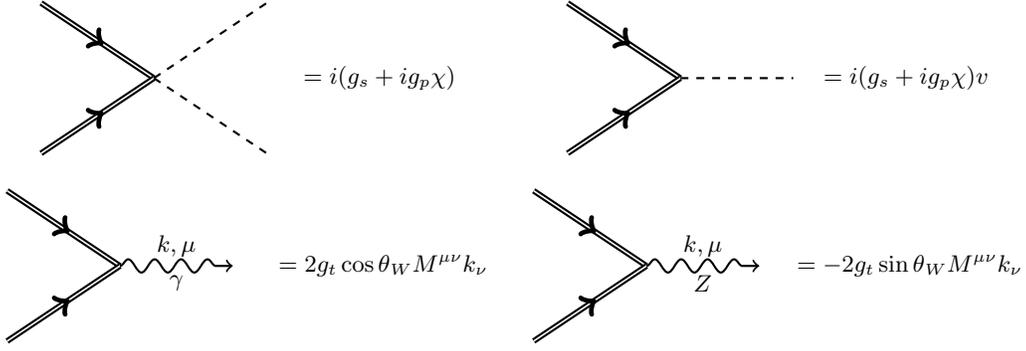

In unitary gauge for the standard model fields, after spontaneous symmetry
breaking and diagonalizing the gauge boson sector we get the following
Lagrangian 
\begin{equation}
\mathcal{L}_{int}= \frac{1}{2}\bar\psi (g_{s} \mathbf{1}+ i g_{p}\chi) \psi
\left(H+ v\right)^2 + g_{t} \cos\theta_{W} \bar\psi M_{\mu\nu} \psi
F^{\mu\nu} - g_{t} \sin\theta_{W} \bar\psi M_{\mu\nu} \psi Z^{\mu\nu} ,
\label{lag}
\end{equation}
where $H$ stands for the Higgs field, $v$ denotes the Higgs vacuum
expectation value and $F^{\mu\nu}, Z^{\mu\nu}$ are the electromagnetic and $%
Z^0$ stress tensors respectively. The Feynman rules arising from the
Lagrangian in Eq. (\ref{lag}) are given in Fig. \ref{FR}.

\begin{figure}[t]
\begin{tikzpicture} 
\draw[dm,thick] (-5.5,1) -- (-4,0) ;
\draw[dm,thick] (-5.5,-1) -- (-4,0) ;
\draw[scalar,thick] (-4,0) -- (-2.5,1) ;
\draw[scalar,thick] (-4,0) --  (-2.5,-1) ;
\node at(-1,0){$=i (g_s  +i g_{p}\chi) $};
\draw[dm,thick] (1.5,1) --   (3,0) ;
\draw[dm,thick] (1.5,-1) --  (3,0) ;
\draw[scalar,thick] (3,0) -- (4.5,0) ;
\node at(6,0){$=i (g_s  +i g_{p}\chi  )v $};
\end{tikzpicture}
\end{figure}

\section{Dark matter with a mass $M<M_{Z}/2$ : $Z^0\to \bar{D}D$ and $H\to 
\bar{D}D$ decays.}


The Lagrangian in Eq.(\ref{lag}) induces transitions between the standard
model and dark sectors. Annihilation of dark matter into standard model
particles such as $\bar{D}D \to \bar{f}f, \gamma\gamma, W^{+}W^{-},
Z^{0}Z^{0}, HH, Z^{0}\gamma, H\gamma, Z^{0} H$ which could be important in
the description of dark matter relic density are induced by these
interactions under appropriate kinematical conditions. Also, for dark matter
mass below half the $Z^0$ mass ($M<M_{Z}/2$), the decays $Z^0\to \bar{D}D$
and $H\to \bar{D}D$ are kinematically permitted and contribute to the
invisible $Z^0$ and $H$ widths respectively. In this work we consider this
mass region and work out the predictions of the formalism for the dark
matter relic density.

\begin{figure}[h!]
\includegraphics[scale=0.5]{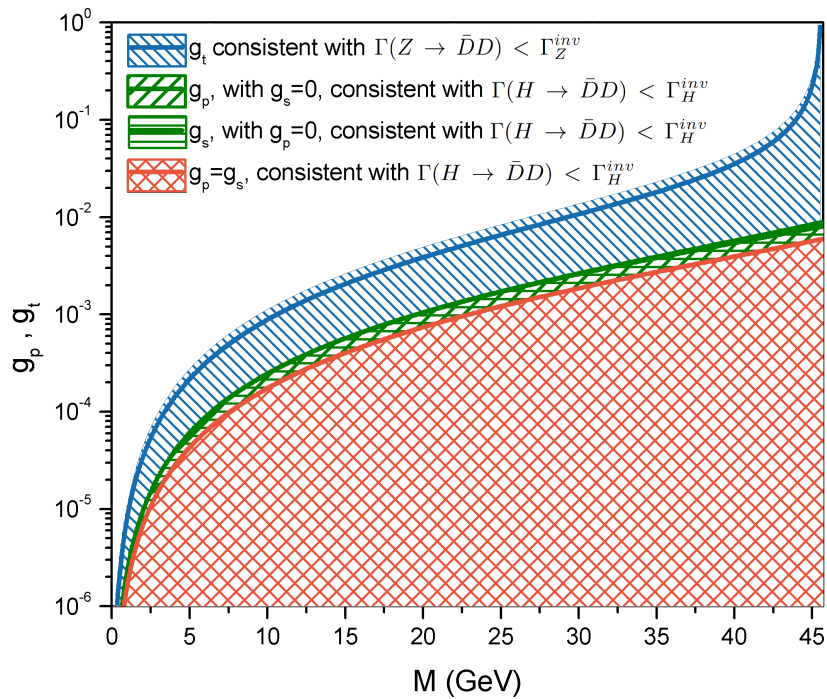} 
\caption{Parameter space for $g_{t}, g_s $ and $g_p$ consistent 
$\Gamma (Z\to \bar{D}D)< \Gamma^{inv}_{Z}=1.4 \pm 1.5~ MeV$ 
and $\Gamma (H\to \bar{D}D)<\Gamma^{inv}_{H}=1.14 \pm 0.04 ~ MeV$ for $M<M_Z/2$. Solid lines correspond to the central values 
of the invisible decay widths. }
\label{gtgsgp}
\end{figure}

A straightforward calculation yields the following invariant amplitude for
the $Z^0(k,\epsilon)\to D(p_1) \bar{D}(p_2)  $ decay 
\begin{equation}
-i\mathcal{M}=2 g_{t}S_{W}\bar{\mathcal{U}}(p_1,\lambda_1)M^{\mu\nu}k_{\nu}%
\mathcal{V}(p_2,\lambda_2)) \epsilon_{\mu} (k),
\end{equation}
where $S_{W}=\sin\theta_W$. The calculation of the average squared amplitude
can be reduced to a trace of products of operators in the covariant basis of 
$(1,0)\oplus(0,1)$ representation space, in a procedure similar to
conventional calculations with Dirac fermions. We obtain
\begin{equation}
|\bar{\mathcal{M}}|^{2}=\frac{4}{3} g^2_{t}S^2_{W}Tr \left[ \frac{S(p_1)+M^2%
}{2M^2}M^{\mu\nu}\frac{S(p_2)+M^2}{2M^2}M^{\alpha\beta} \right]
k_{\nu}k_{\beta}(-g_{\mu\alpha} + \frac{k_\mu k_\alpha}{M^2_Z}).
\end{equation}
The trace-ology of matrices in $(1,0)\oplus (0,1)$ space is deferred to an appendix. 
Using results in the appendix we obtain the corresponding decay width as  
\begin{equation}
\Gamma(Z^{0}\to \bar{D}D) =\frac{g_{t}^2 S^2_W}{24 \pi M^4} (M_Z^2-4
M^2)^{3/2} (M_Z^2+2 M^2 ).
\end{equation}
The invisible width $\Gamma^{inv}_{exp}(Z)=499.0\pm 1.5 ~MeV$ reported by the
Particle Data Group \cite{Patrignani:2016xqp}, includes the decay to $\nu 
\bar{\nu}$. We use the SM prediction for the latter 
\begin{equation}
\Gamma_{SM}(Z^{0}\to \bar{\nu}\nu)\equiv \sum_{i} \Gamma_{SM}(Z^{0}\to \bar{%
\nu}_{i}\nu_{i})= \sum_{i,\alpha} U_{i\alpha}^{2}\frac{M^2_{Z}}{24\pi v^{2}}%
\sqrt{M^2_{Z}-4 m^2_{\nu_{i}}} =\frac{M^3_{Z}}{8\pi v^{2}} =\frac{\sqrt{2}%
G_{F}M^3_{Z}}{8\pi }.
\end{equation}
where in the last step we neglected the neutrino masses and used the
unitarity of the PMNS matrix elements. The Particle Data Group report the
value $M_{Z}=91.1876 \pm 0.0021 ~GeV$ while the $\mu-Lan$ collaboration
reported the most precise measurement of the Fermi constant as $%
G_F=1.1663788(6)\times 10^{-5} GeV^{-2}$ \cite{Webber:2010zf} . Using these
values we get 
\begin{equation}
\Gamma_{SM}(Z^{0}\to \bar{\nu}\nu)= 497.64\pm0.03 ~MeV.
\end{equation}
Subtracting this quantity from the PDG reported value for the invisible
width we get the constraint $\Gamma (Z\to \bar{D}D)< \Gamma^{inv}_{Z}\equiv
\Gamma^{inv}_{exp}(Z)-\Gamma_{SM} (Z\to \bar{\nu}\nu)=1.4 \pm 1.5~MeV$. This
width depends on the coupling $g_{t}$ and the dark matter mass $M$, hence
the invisible $Z^0$ width constrain these parameters to the region shown in
Fig. \ref{gtgsgp}. 

Similar calculations for the $H\to \bar{D}D$ decay yield the following decay
width  
\begin{equation}
\Gamma(H\to \bar{D}D) = \frac{v^2}{32\pi M^2_{H} M^4} \sqrt{ M_H^2-4 M^2} %
\left[g_s^2 \left(M_H^2\left( M_H^2-4 M^2\right) +6 M^4\right) + g_p^2 M_H^2
\left(M_H^2-4 M^2\right)\right],
\end{equation}
The $H\to \bar{D}D$ width depends on the unknown $g_s$, $g_p$ couplings and
on the dark matter mass. This channel contributes to the invisible Higgs
width which has been recently reported in 
\cite{Patrignani:2016xqp,Khachatryan:2016whc} as 
$\Gamma^{inv}_{H}=1.14 \pm 0.04 ~MeV$. In this case, the contribution of the $\nu \bar{\nu}$ channel is
negligible. The constraints on $g_{s},g_{p}$ arising from the 
$\Gamma (H\to \bar{D}D)<\Gamma^{inv}_{H}$ condition are also shown in Fig. \ref{gtgsgp}. 
The solid lines correspond to the central values and the shadow regions to the one sigma regions. 
We conclude from this plot that the coupling of the spin portal $g_{t}$ in
general can be larger than those of the Higgs portal $g_s$ or $g_p$, by at
least one order of magnitude. 


\section{Dark matter relic density.}


\subsection{Boltzman equation.}

The evolution of the dark matter comoving number density $n_{D}(T)$ is described by the Boltzmann equation \cite{Dodelson:2003ft} 
\begin{equation}
\frac{dY}{dx}=- \frac{\lambda (x)}{ x^2}(Y^2-Y^2_{eq}),  
\label{Bolt}
\end{equation}
where $x=M/T$, $Y(x)=n_{D}(x)/T^3$ and 
\begin{equation}
\lambda (x)\equiv \frac{M^3\langle \sigma v_r \rangle}{H(M) }.
\label{lambda}
\end{equation}
Here, $H(M)=M^2\sqrt{\frac{8\pi^3G_{N} g_{*}(M)}{90}} $ stands for the Hubble
parameter at the dark mass scale, $M$, with  $G_{N}=6.70861(31)\times
10^{-39} GeV^{-2}$ denoting the Newton gravitational constant \cite%
{Patrignani:2016xqp}, $g^{*}(M)$  standing for the relativistic effective
degrees of freedom at $T=M$ in the thermal bath and 
\begin{equation}
Y_{eq}(x)=\frac{n_{D}^{eq}}{T^3}=\frac{g_{D}}{T^3}\int \frac{d^3p}{(2\pi)^3}%
\frac{1}{e^{\frac{E}{T}}-1} = \frac{3}{2\pi^2}\int_x^{\infty}\frac{u\sqrt{%
u^2-x^2} du}{e^{u}-1} \approx \frac{3}{2\pi^2}\int_x^{\infty}e^{-u}u\sqrt{%
u^2-x^2} du .
\label{Yeq}
\end{equation}
The thermal average $\langle \sigma v_r \rangle$ includes all channels for
the annihilation $D(p_1)\bar{D}(p_2)\to X (p_3) Y(p_4)$ of dark matter into
standard model particles $X,Y$ in the thermal bath and it is given by 
\begin{equation}
\langle \sigma v_r \rangle =\frac{1}{n_D^{eq} n_{\bar{D}}^{eq}} \int \frac{%
g_D d^3 p_1}{(2\pi)^2} e^{-E_1 /T} \int \frac{g_{\bar{D}} d^3 p_2}{(2\pi)^2}
e^{-E_2/T} \sigma v_r,
\end{equation}
where $g_{D} $ ($g_{\bar{D}} $)denotes the number of internal d.o.f of the
dark matter particle (antiparticle), $v_{r}$ stands for the dark matter
particle-antiparticle relative velocity and $\sigma$ is the conventional
cross section for the $D(p_1)\bar{D}(p_2)\to X (p_3) Y(p_4)$ process.

A qualitative analysis of the solution of Eq. (\ref{Bolt}) assuming the
freezing of dark matter at some temperature which would explain dark matter
relic density, shows that dark matter must be non-relativistic at the time
of its decoupling from the cosmic plasma \cite{Dodelson:2003ft}. This is
consistent with data on dark matter relic density extracted from precision
measurement of the cosmic background radiation \cite%
{Ade:2015xua,Patrignani:2016xqp}. In this case, it is a good approximation
to perform a non-relativistic expansion of $\langle \sigma v_r \rangle$
keeping only the leading terms in the expansion in powers of $v_r<<1$. This
expansion requires the calculation of the flux for dark matter particles in
the thermal bath, which can be written as \cite{Gondolo:1990dk, Cannoni:2016hro} 
\begin{equation}
F=4\sqrt{(p_1\cdot p_2)^2-M^4}=2(s-M^2)v_{r}  \label{flux}
\end{equation}
where $v_{r}$ is related to $s$ as 
\begin{equation}
s= 2M^2\left(1+\frac{1}{\sqrt{1-v^2_{r} }} \right)=4M^2 + M^2 v^2_r+... .
\end{equation}
In the last step we performed the non-relativistic expansion for $v_r<<1$.
The cross section $\sigma$ is a function of $s$ thus using Eq.(\ref{flux})
the leading terms in the expansion are 
\begin{equation}
\sigma v_r = a +b v^2_{r},
\end{equation}
and performing the thermal average we obtain 
\begin{equation}
\langle \sigma v_r \rangle = a +\frac{6b}{x}.  
\label{sigmav}
\end{equation}

For non-relativistic dark matter with $M<M_{Z}/2$,  the kinematically allowed 
channels are $\bar{D}D\to \bar{f}f$ for fermions with $m_{f}<M$ and 
$\bar{D}D\to \gamma\gamma$. In the
following we calculate the corresponding cross sections in our formalism, perform the
non-relativistic expansion and work out the predictions for
the $a, b$ coefficients.

\subsection{Annihilation of dark matter into a fermion-antifermion pair.}

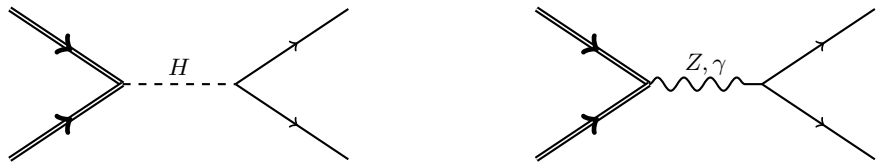
\begin{figure}[t]
\centering
\begin{tikzpicture}
\draw[dm,thick] (-5.5,1) --   (-4,0) ;
\draw[dm,thick] (-5.5,-1) --  (-4,0) ;
\draw[scalar,thick] (-4,0) -- node[above]{$H$}  (-2.5,0) ;
\draw[fermion,thick] (-2.5,0) -- (-1,1) ;
\draw[fermion,thick] (-2.5,0) -- (-1,-1) ;
\draw[dm,thick] (1.5,1) --   (3,0) ;
\draw[dm,thick] (1.5,-1) --  (3,0) ;
\draw[gauge,thick] (3,0) -- node[above]{$Z, \gamma$} (4.5,0) ;
\draw[fermion,thick] (4.5,0) -- (6,1) ;
\draw[fermion,thick] (4.5,0) -- (6,-1) ;
\end{tikzpicture}
\caption{Feynman diagrams for $\bar{D} D\to \bar{f}f$. }
\label{DDff}
\end{figure}

There are three contributions to the process $D(p_1)\bar{D}(p_2) \rightarrow
f(p_3)\bar{f}(p_4) $ shown in Fig. \ref{DDff}. 
The corresponding amplitudes are given by 
\begin{eqnarray}
-i\mathcal{M}_{H} &=&i\frac{m_{f}}{s-M_{H}^{2}}\bar{u}\left( p_{3}\right)
v\left( p_{4}\right) \mathcal{\bar{V}}\left( p_{2}\right) \left(
g_{s}I+ig_{p}\chi\right) \mathcal{U}\left( p_{1}\right) ,  \notag \\
-i\mathcal{M}_{\gamma} &=&-\frac{4Q_{f} g_{t}M_{W} S_{W} C_{W} }{v s}\bar{u}%
\left( p_{3}\right) \gamma^{\mu}v\left( p_{4}\right) \mathcal{\bar{V}}\left(
p_{2}\right) M_{\mu\beta}\left( p_{1}+p_{2}\right)^{\beta}\mathcal{U}\left(
p_{1}\right) , \\
-i\mathcal{M}_{Z} &= & \frac{g_{t}M_{Z} S_{W} }{v (s-M_{Z}^{2})}\bar{u}%
\left( p_{3}\right) \gamma^{\mu}\left( A_{f}+B_{f}\gamma_{5}\right) v\left(
p_{4}\right) \mathcal{\bar{V}}\left( p_{2}\right) M_{\mu\beta}
\left(p_{1}+p_{2}\right){\beta} \mathcal{U}\left( p_{1}\right).  \notag
\end{eqnarray}
Here, $C_{W}=Cos\theta_{W}$, $Q_{f}$ stands for the fermion charge in units
of the proton charge $e$, while the $A_{f}, B_{f}$ factors are related to the
corresponding fermion weak isospin $T^{f}_3$ as 
\begin{equation}
A_{f}=2 T^{f}_3- 4 Q_{f} S^{2}_{W}, \qquad B_{f}= -2 T^{f}_3.
\end{equation}

A straightforward calculation yields the following average squared amplitude
in terms of the Mandelstam variables 
\begin{align}
\left\vert \overline{\mathcal{M}_{\bar{f}f} }\right\vert ^{2}= & -\frac{%
g_{t}^2 M_Z^2 S_W^2}{9 M^4 v^2 \left(s-M_Z^2\right)^2} \left[4 M^2
\left(A_{f}^2+B_{f}^2\right) m_f^4 \left(4 M^2-s\right) \right.  \notag \\
& + 4 m_f^2 \left. \left(4 M^2-s\right) \left(A_{f}^2 M^2 \left(2
M^2+s-t-u\right)+B_{f}^2 \left(2 M^4-M^2 (s+t+u)-s^2\right)\right) \right. 
\notag \\
& + \left. \left(A_{f}^2+B_{f}^2\right) \left(16 M^8-4 M^6 (s+4 (t+u))+4 M^4 (t+u)
(s+t+u)+M^2 \left(4 s^3-2 s \left(t^2+u^2\right)\right)+s^2
\left((t-u)^2-s^2\right)\right)\right]  \notag \\
&+ \frac{8 A_{f} C_W Q_f g_{t}^2 M_W M_Z S_W^2 }{9 M^4 s v^2 \left(s-M_Z^2\right)}%
\left[4 M^2 m_f^2 \left(4 M^2-s\right) \left(2 M^2+s-t-u\right)+4 m_f^4
\left(4 M^4-M^2 s\right)+16M^8 \right.  \notag \\
& \left. -4 M^6 (s+4 (t+u))+4 M^4 (t+u) (s+t+u)+M^2 \left(4 s^3-2 s
\left(t^2+u^2\right)\right)+s^2 \left((t-u)^2-s^2\right)\right]  \notag \\
& + \frac{ A_{f} m_f^2 g_s g_{t} M_Z S_W }{9 M^4 v \left(s-M_H^2\right)
\left(s-M_Z^2\right)} s \left(2M^2-s\right)(t-u)  \notag \\
& - \frac{16 C_W m_f^2 Q_f g_s g_{t} M_W S_W }{9 M^4 v \left(s-m_H^2\right)}
\left(2 M^2-s\right) (t-u)  \notag \\
& - \frac{16 C_W^2 Q_f^2 g_{t}^2 M_W^2 S_W^2}{9 M^4 s^2 v^2} \left[4 M^2 m_f^2
\left(4 M^2-s\right) \left(2 M^2+s-t-u\right)+4 m_f^4 \left(4 M^4-M^2
s\right) \right.  \notag \\
& \left. +16 M^8-4 M^6 (s+4 (t+u))+4 M^4 (t+u) (s+t+u)+M^2 \left(4 s^3-2 s
\left(t^2+u^2\right)\right)+s^2 \left((t-u)^2-s^2\right)\right]  \notag \\
& + \frac{m_f^2}{9 M^4 \left(s-M_H^2\right)^2} \left(s-4 m_f^2\right) \left[
g_p^2 s \left(s-4 M^2\right)+g_s^2 \left(6 M^4-4 M^2 s+s^2\right)\right] .
\end{align}
Integrating the final state phase space finally we obtain the following
cross section for $\bar{D}D\to \bar{f}f$ where we can easily identify the
individual contributions from $H, Z^0$ and $\gamma$ exchange as well as the $%
Z^{0}-\gamma$ interference: 
\begin{align}
\sigma_{\bar{f}f} (s)& =\frac{1}{ 72\pi M^4 \sqrt{s}} \frac{\sqrt{%
s-4m_{f}^{2}}}{F} \left[ \frac{ m_f^2 \left(s-4 m_f^2\right) \left( g_p^2 s
\left(s-4 M^2\right)+g_s^2 \left(6 M^4-4 M^2 s+s^2\right)\right)}{
\left(s-M_H^2\right)^2} \right.  \notag \\
&\left. +\frac{2 g_{t}^2 M_Z^2 S_W^2 s\left(s-4 M^2 \right) \left(2
M^2+s\right) \left(2 \left(A_{f}^2-2 B_{f}^2\right) m_f^2+s
\left(A_{f}^2+B_{f}^2\right)\right)}{3 v^2 \left(s-M_Z^2\right){}^2} \right.  \notag
\\
&\left. +\frac{32 C_W^2 Q_f^2 g_{t}^2 M_W^2 S_W^2 \left(s-4 M^2 \right)
\left(2 M^2+s\right) \left(2 m_f^2+s\right)}{3 v^2 s} \right.  \notag \\
&\left. -\frac{16 A_{f} C_W Q_f g_{t}^2 M_W M_Z S_W^2 \left(s-4 M^2 \right)
\left(2 M^2+s\right) \left(2 m_f^2+s\right)}{3 v^2 \left(s-M_Z^2\right)} %
\right].
\end{align}
Notice that the $H-Z$ and $H-\gamma$ interferences vanish after integration of 
phase space. 

\subsection{ Dark matter annihilation into two photons}

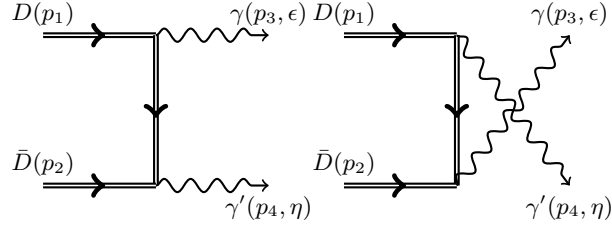
\begin{figure}[t]
\centering
\begin{tikzpicture}
\draw[dm,thick] (-6.5,1) node[above]{$D (p_1)$} --   (-5,1) ;
\draw[dm,thick] (-5,1)  --   (-5,-1) ;
\draw[dm,thick] (-6.5,-1) node[above]{$\bar{D} (p_2)$} --  (-5,-1) ;
\draw[->, gauge,thick] (-5,1) -- (-3.5,1) node[above]{$\gamma (p_3, \epsilon)$};
\draw[->, gauge,thick] (-5,-1) -- (-3.5,-1) node[below]{$\gamma^{\prime}(p_4,\eta)$};
\draw[dm,thick] (-2.5,1)  node[above]{$D (p_1)$}  --   (-1,1) ;
\draw[dm,thick] (-1,1) --   (-1,-1) ;
\draw[dm,thick] (-2.5,-1) node[above]{$\bar{D} (p_2)$} --  (-1,-1) ;
\draw[->, gauge,thick] (-1,1) -- (0.5,-1) node[below]{$\gamma^{\prime}(p_4,\eta)$};
\draw[->, gauge,thick] (-1,-1) -- (0.5,1) node[above]{$\gamma (p_3, \epsilon)$};

\end{tikzpicture}
\caption{Feynman diagrams for $\bar{D}D\rightarrow \protect\gamma \protect\gamma $. 
}
\label{2g}
\end{figure}

This process is induced by the $t$ and $u$ channel dark matter exchange
shown in Fig. \ref{2g}. The corresponding amplitudes are given by 
\begin{eqnarray}
-i\mathcal{M}_{t} &=& i\frac{2g^{2}_{t}C^{2}_{W}}{M^{2}} \bar{V}%
(p_{2},\lambda_{2})M_{\alpha\beta} \frac{S(p_{1}-p_{3})-t+2M^{2}}{t-M^{2}}
M_{\mu\nu} U(p_{1},\lambda_{1}) p_{4}^{\alpha} \eta^{\beta}( p_{4})
p_{3}^{\mu} \epsilon^{\nu}( p_{3}) , \\
-i\mathcal{M}_{u} &=& i\frac{2g^{2}_{t}C^{2}_{W}}{M^{2}} \bar{V}%
(p_{2},\lambda_{2})M_{\mu\nu} \frac{S(p_{1}-p_{4})-u+2M^{2}}{u-M^{2}}
M_{\alpha\beta} U(p_{1},\lambda_{1}) p_{4}^{\alpha} \eta^{\beta}( p_{4})
p_{3}^{\mu} \epsilon^{\nu}( p_{3}).
\end{eqnarray}
The average squared amplitude is given by 
\begin{equation}
|\overline{\mathcal{M}_{\gamma\gamma}}|^{2}=\left( \frac{2g^{2}_{t}C^{2}_{W}%
}{3 M^{2}} \right)^{2} Tr \left[ \frac{S(p_{2}) + M^{2}}{2M^{2}}
T_{\alpha\beta\mu\nu} \frac{S(p_{1})+ M^{2}}{2M^{2}} \bar{T}%
_{\sigma~\rho}^{~\beta~\nu} \right] p_{3}^{\mu} p_{3}^{\rho}
p_{4}^{\alpha}p_{4}^{\sigma},
\end{equation}
where 
\begin{eqnarray}
T_{\alpha\beta\mu\nu} &=&M_{\alpha\beta} \frac{S(p_{1}-p_{3})-t+2M^{2}}{%
t-M^{2}} M_{\mu\nu} + M_{\mu\nu} \frac{S(p_{1}-p_{4})-u+2M^{2}}{u-M^{2}}
M_{\alpha\beta} , \\
\bar{T}_{\alpha\beta\mu\nu} &=&M_{\mu\nu} \frac{S(p_{1}-p_{3})-t+2M^{2}}{%
t-M^{2}} M_{\alpha\beta} + M_{\alpha\beta} \frac{S(p_{1}-p_{4})-u+2M^{2}}{%
u-M^{2}} M_{\mu\nu} .
\end{eqnarray}

A straightforward calculation using the algebraic relations in the appendix yields 
\begin{align}
\overline{|\mathcal{M}_{\gamma \gamma }|^{2}}& =\frac{2C_{W}^{4}g_{t}^{4}}{%
9M^{8}\left( t-M^{2}\right) ^{2}\left( u-M^{2}\right) ^{2}}\left[ 6\left(
tu\right) ^{4}+2\left( tu\right) ^{3}\left( -13M^{4}+11M^{2}s+2s^{2}\right)
\smallskip \right.   \notag \\
& +\left( tu\right) ^{2}\left(
42M^{8}-76M^{6}s+33M^{4}s^{2}+4M^{2}s^{3}+2s^{4}\right) \smallskip   \notag
\\
& +2M^{2}tu\left(
-15M^{10}+43M^{8}s-44M^{6}s^{2}+17M^{4}s^{3}-6M^{2}s^{4}+2s^{5}\right)
\smallskip   \notag \\
& +\left. M^{4}\left(
8M^{12}-32M^{10}s+51M^{8}s^{2}-40M^{6}s^{3}+25M^{4}s^{4}-12M^{2}s^{5}+2s^{6}%
\right) \smallskip \right] 
\end{align}

Integrating the final state phase space we get the following cross section 
\begin{align}
\sigma_{\gamma\gamma}(s)&=\frac{1}{F\sqrt{1-\frac{4 M^2}{s}}} \frac{C_W^4
g_{t}^4 }{540 \pi M^8 } \left[120 M^4 \left(4 M^4-3 M^2 s-2 s^2\right)
\tanh^{-1}\sqrt{1-\frac{4 M^2}{s}} \right.  \notag \\
&\left. + s \sqrt{1-\frac{4 M^2}{s}} \left(-10 M^6+228 M^4 s-99 M^2 s^2+43
s^3\right) \right] .
\end{align}

\subsection{Dark matter relic density}
Expanding the $\bar{D}D\to \bar{f}f$ and $\bar{D}D\to \gamma\gamma$ cross
sections we get 
\begin{equation}
\sigma v_{r}\equiv \sigma_{\gamma\gamma} v_{r}+ \sum_{f}\sigma_{\bar{f}%
f}v_{r} =a +b v_{r}^2
\end{equation}
where the sum runs over all the kinematically allowed fermion states ($%
m_{f}<M$) and 
\begin{eqnarray} \label{ab}
a&=&\frac{29 C_W^4 g_{t}^4}{18 \pi M^2} + \sum_{f} \frac{ N_f g_s^2 m_f^2
\left(M^2-m_f^2\right)^{\frac{3}{2}} }{12 \pi M^3 \left(M_H^2- 4
M^2\right){}^2}  \nonumber  \\
b&=& \frac{365 C_W^4 g_{t}^4}{216 \pi M^2} + \sum_{f} \frac{N_f \sqrt{M^2-m_f^2%
}}{864 \pi M^5} \left(\frac{96 M^4 g_{t}^2 M_Z^2 S_W^2 \left(\left(A_{f}^2-2
B_{f}^2\right) m_f^2 +2 M^2\left(A_{f}^2+B_{f}^2\right)\right)}{v^2 \left(M_Z^2-4
M^2\right){}^2} \right. \\
&& \left. +\frac{192 A_{f} M^2 C_W Q_f g_{t}^2 M_W M_Z S_W^2 \left(m_f^2+2
M^2\right)}{v^2 \left(M_Z^2-4 M^2\right)} +\frac{96 C_W^2 Q_f^2 g_{t}^2 M_W^2
S_W^2 \left(m_f^2+2 M^2\right)}{v^2} \right.  \nonumber  \\
&& \left. -\frac{6 M^2 m_f^2 \left(8 g_p^2 \left(4 M^2-M_H^2\right)
\left(M^2-m_f^2\right)+g_s^2 \left(-8 m_f^2 \left(M^2-M_H^2\right)-11 M^2
M_H^2+20 M^4\right)\right)}{\left(M_H^2-4 M^2\right)^3} \right. \nonumber  \\
&& \left. -\frac{9 M^2 m_f^2 g_s^2 \left(4 M^2-5 m_f^2\right)}{\left(M_H^2-4
M^2\right){}^2}\right) ,  \nonumber 
\end{eqnarray}
with $N_f=3$ for quarks and $N_f=1$ for leptons. We can see in these
equations that for the mass region $M<M_{Z}/2$ the Higgs portal
contributions are suppressed compared to the spin portal ones by factors $%
m^2_f/M^2_H$.

\begin{figure}[h]
\includegraphics[scale=0.4]{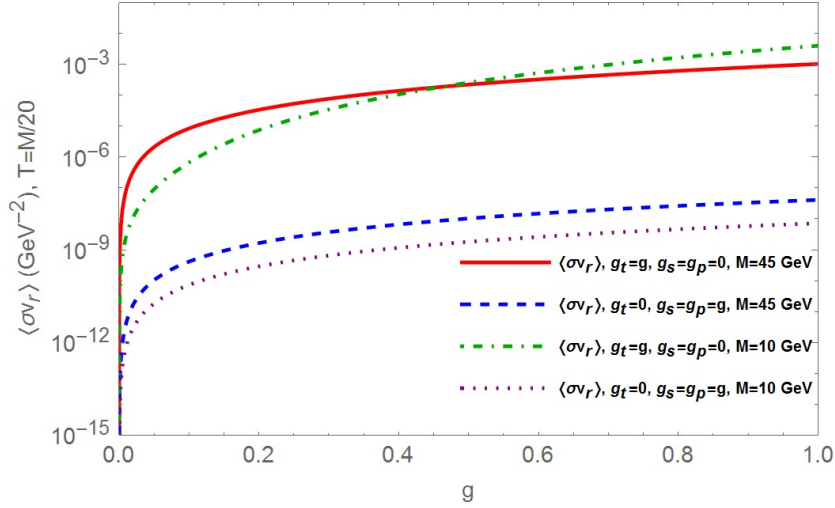}
\caption{Individual contributions of the spin portal ($g_t=g,~g_{s}=g_{p}=0$) and the Higgs
portal ($g_t=0,~g_{s}=g_{p}=g$) to $\langle \protect\sigma v_r \rangle$. Similar results are 
obtained in the second case when varying independently $g_{s}$ or $g_{p}$. }
\label{fig:HiggsSpinPortal}
\end{figure}

In Fig. (\ref{fig:HiggsSpinPortal}) we analize the Higgs and spin portal
contributions to $\langle \sigma v_r \rangle$ as a function of the couplings
for different values of the dark matter mass. In general, we find that Higgs
portal contributions are negligible compared to the contributions of the
spin portal. Therefore, we will neglect the contribution of the Higgs portal
for the calculation of the relic density in the following.

\begin{figure}[h]
\includegraphics[scale=0.4]{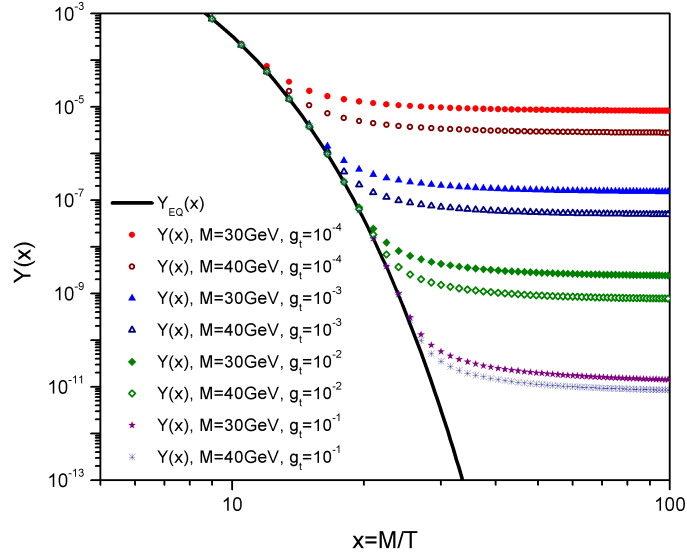}
\caption{Solution of the Boltzman equation for different values of $M$ and $g_{t}$ showing that dark 
matter decouples in the non-relativistic regime. The solid line corresponds to $Y_{eq}(x)$. }
\label{solbolteq}
\end{figure}

Using Eqs. (\ref{sigmav},\ref{ab}), we numerically solve Boltzman equation (\ref{Bolt})
for different values of $g_{t}$ and $M$, matching the solution $Y(x)$ with the 
equilibrium solution $Y_{eq}(x)$ in Eq.(\ref{Yeq}) at high temperatures, i.e., in the relativistic regime $x<<1$. 
In Fig.(\ref{solbolteq}) we show the solutions for some specific values of $g_{t}$ and $M$. Clearly, at some
temprature $T_f$ the solution $Y(x)$ departs from the equilibrium solution 
$Y_{eq}(x)$ and dark matter decouples from the cosmic plasma in the non-relativistic regime,  $x>>1$. 

In order to find the dark matter relic density we need to calculate $Y$ for the present temperature $T_{0}$. 
This can be done from the numeric solution to Boltzman equation for specific values of $g_{t}$ and $M$ 
scanning the parameter space consistent with the measured relic density. It is however more illustrative to 
follow the semi-analytic procedure that takes advance of the freezing mechanism. 
For $x>x_f$ we have $Y(x)>>Y_{eq}(x)$ an we can find an approximate solution
neglecting $Y_{eq}(x)$ in the r.h.s of Eq.(\ref{Bolt}) and integrating from $T_f$ to a given temperature $T$, 
which for our purposes we take as the present temperature $T_0$, to obtain 
\begin{equation}
\frac{1}{Y(x_0)}=\frac{1}{Y(x_f)}+\sqrt{ \frac{90} {8\pi^{3}G_{N}}}M
\int_{x_f}^{x_0} \frac{ \langle \sigma v \rangle}{\sqrt{g_{*}(x)}x^2}dx.
\label{solY}
\end{equation}
The relic dark matter density is given by 
\begin{equation}
\Omega_{DM}=\frac{\rho_{DM}(x_0)}{\rho_c}=\frac{(n_{D}(x_0)+n_{\bar{D}%
}(x_0))M}{\rho_c} =\frac{2n_{D}(x_0)M}{\rho_c}=\frac{2MY(x_0)T^{3}_{0}}{%
\rho_c},
\end{equation}
where we used $n_{\bar{D}}=n_{D}$ and 
$\rho_{c}= \frac{3H^2_0}{8\pi G_N}=1.05371(5) \times 10^{-5} h^2 GeV/cm^3=8.09619(38) \times 10^{-47} h^2 GeV^4$ 
is the critical density \cite{Patrignani:2016xqp}. Neglecting the term 
$Y(x_f)^{-1}$ in Eq. (\ref{solY}) which turns out to be small compared with
the second term we get 
\begin{equation}
\Omega_{DM} h^2 = \frac{2 T^{3}_{0}h^2}{\rho_c}\sqrt{\frac{8\pi^3 G_{N}}{90}}
\left( \int_{x_f}^{x_0} \frac{ \langle \sigma v \rangle}{\sqrt{g_{*}(x)}x^2}%
dx \right)^{-1} = 4.337\times10^{-11} GeV^{-2} \left( \int_{x_f}^{x_0} \frac{
\langle \sigma v \rangle}{\sqrt{g_{*}(x)}x^2}dx \right)^{-1} 
\label{relic}
\end{equation}
where we used $T_{0}=2.7255(6) K=2.34865(52)\times 10^{-13}GeV$ \cite{Patrignani:2016xqp}. 
Notice that the r.h.s. of this equation depends
on $g_{t}$ and $M$. For a given $M$ we can find the values of $g_{t}$
consistent with the measured value of the relic density. In our calculations we use
the complete function $g_{*}(x)$ but our results are quite similar if we use the average 
over the range of energies considered, $\bar{g}_{*}= 33$.

\begin{figure}[h]
\includegraphics[scale=0.4]{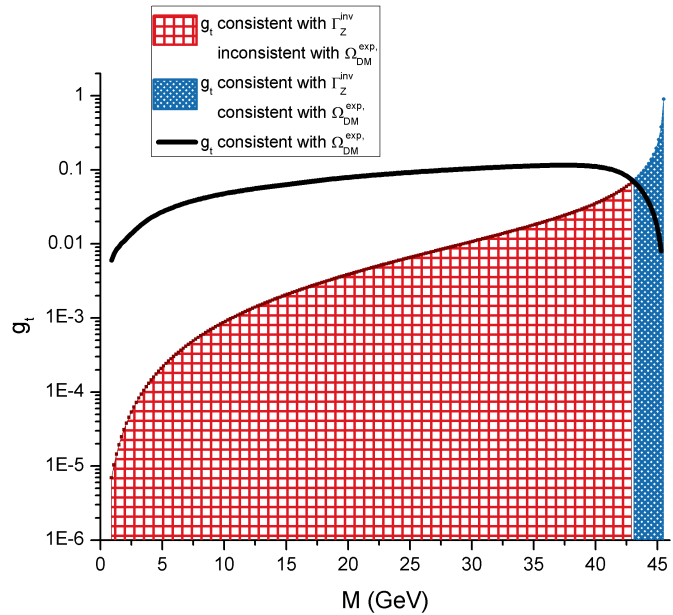}
\caption{Values of $g_{t}$ and $M$ consistent with the
measured dark matter relic density, $\Omega^{exp}_{DM} h^2 =0.1186\pm0.0020 $ (solid line). 
The shadowed region denotes the values consistent with the constraint 
$\Gamma(Z^{0}\to \bar{D}D)<\Gamma^{inv}_{Z}=1.4 \pm 1.5~ MeV$. These constraints exclude masses 
below $43~ GeV$ for dark matter with a $(1,0)\oplus (0,1)$ space-time structure.}
\label{fig:coupCOMP}\centering
\end{figure}

The freezing value $x_{f}$ can be found from the condition that the
annihilation rate equals the expansion rate of the universe  
\begin{equation}
n_{eq}(x_f) \langle \sigma v \rangle (x_f)=H(x_f),
\end{equation}
which using the non-relativistic form for $n_{eq}(x)$ and Eq. (\ref{sigmav})
leads to  
\begin{equation}
\left(a+\frac{6b}{ x_f}\right)\sqrt{x_f}e^{-x_f}=\frac{(2\pi)^3 }{3M } \sqrt{%
\frac{G_{N}g^*(x_f)}{90}}.
\label{freezing}
\end{equation}
The value of $x_{f}$ depends also on $g_{t}$ and $M$, so we have two conditions, 
Eqs. (\ref{relic},\ref{freezing}), for the three variables $x_{f}, g_{t}, M$ which are solved 
numerically to obtain the set of values $g_{t}(M)$ consistent with the measured dark 
matter relic density. The set of values $g_{t}(M)$ is shown in Figure \ref{fig:coupCOMP}. 
We checked also that these solutions are consistent with 
the approximations used, i.e. that decoupling occurs when dark matter is non-relativistic.
The values of $x_{f}$ corresponding to $g_{t}(M)$ lie in the range $23.8<x_f<27.9$, thus $x_{f}>>1$. 
Finally, we directly calculate $Y(x)$ from the numeric general solution of the Boltzman equation 
for the set of values $g_{t}(M)$, matching the solution with $Y_{eq}(x)$ for $x<<x_{f}$ finding indeed  
that $1/Y(x_f)$ is small compared to $1/Y(x_0)$ in Eq.(\ref{solY}). 

Our results are summarized in Figure \ref{fig:coupCOMP}, where it is clear that 
taking into account constraints from the data on the $Z^0$ invisible width and from 
the measured dark matter relic density, dark matter with a $(1,0)\oplus(0,1)$ space-time 
structure must have a mass $M>43 ~GeV$.  

\section{Conclusions and perspectives}

Effective theories for the interaction of dark matter with standard model
fields has been done mainly assuming space-time structures for dark matter
similar to those of the standard model fields, i.e., dark matter fields transforming 
in the $(0,0)$, $(\frac{1}{2},0)\oplus (0,\frac{1}{2})$ or $(\frac{1}{2}, \frac{1}{2})$ 
representations of the HLG.

In this work we study the possibility of a $(1,0)\oplus(0,1)$ space-time structure for dark
matter fields. Assuming that dark matter fields are standard model singlets, we find 
three lowest order terms which are dimension-four in the corresponding effective theory. 
Two of them couple  the Higgs to dark matter and the third one couples the photon 
and $Z^0$ fields to higher multipoles of the spin-one dark matter fields, yielding 
a \textit{spin portal} to dark matter.

We start the study of the phenomenology derived from our 
proposal considering dark matter mass $M<M_Z/2$, in whose case 
the $H\to \bar{D}D$ and $Z^0\to \bar{D}D$
are kinematically permitted and contribute to the Higgs and $Z^0$ invisible
decay widths. We use experimental results on these widths to put upper
limits to the corresponding low energy constants. In general we find
stringent constraints for the couplings of the Higgs portal: 
$g_s,g_p \leq 10^{-3}$ and less stringent constraints on the spin portal coupling $g_{t}$.

For dark matter mass in this region, non-relativistic dark matter can annihilate into a photon 
pair or into a fermion-anti-fermion pair if $M>m_f$. We calculate these processes 
in our formalism and use them to calculate the corresponding dark matter relic 
density.  We find that the contribution of the Higgs portal to the dark matter
relic density is negligible and the main contribution comes from the spin portal. 
Taking into account the constraints from the $Z^0$ invisible width, we find that 
a proper description of the measured dark matter relic density imposes the 
lower bound $M>43 ~ GeV$ for dark matter with a $(1,0)\oplus(0,1)$ space-time 
structure.

The spin portal yields a new avenue for the possible transitions between the dark matter and  
standard model sectors whose phenomenological consequences  are worthy to 
explore further. Here, we study the low mass 
regime, $M<M_{Z}/2$, where low energy constants can be constrained from the $H$ and $Z^0$ 
invisible widths. For $M>M_{Z}/2$, the $Z^{0}\to\bar{D}D$ decay is 
kinematically forbidden and we loose the corresponding constraints on $g_{t}$. 
Furthermore, in this regime, depending on the kinematics, new channels for the 
annihilation of dark matter such as $\bar{D}D \to Z^{0}\gamma, H\gamma, W^{+}W^{-},
Z^{0}Z^{0}, Z^{0} H, HH, \bar{t}t$ open and must be 
considered in the analysis of the dark matter relic density. On the other hand, 
some experiments of direct detection of dark matter attempt to detect nuclear recoil due to the
scattering of nuclei with dark matter, ultimately related to the quark-dark matter 
scattering, which takes place in our formalism. It is important to calculate these effects in order to 
further constrain the possible values of the mass and couplings of spin-one dark matter. 
Finally, it would be important to study 
all processes involving dark matter so far analyzed at the LHC on the light of 
spin-one dark matter fields.

\acknowledgments
 Work supported by CONACyT M\'{e}xico under project CB-259228. H.H.A. acknowledges 
 CONACyT for a scholarship and DAIP-UG for a grant 
 under the Call for Support to Graduate Studies 2017. 

\section{Appendix: Trace-ology for $(1,0)\oplus(0,1)$.}

In this appendix we collect the trace relations necessary for the
calculations in this work. The covariant basis for the $(1,0)\oplus(0,1)$
representation space is given by the set of $6\times 6$ matrices $\{ \mathbf{%
1}, \chi, S^{\mu\nu}, \chi S^{\mu\nu}, M^{\mu\nu}, C^{\mu\nu\alpha\beta } \}$
where $\mathbf{1}$ is the identity matrix. The first principles construction
of these matrices can be found in \cite{Gomez-Avila:2013qaa} and their
explicit form depends on the basis chosen for the states in the $%
(1,0)\oplus(0,1)$ representation. All the calculations in this work are
representation independent and rely only on their algebraic properties. The
starting point are first principles construction of the rest-frame parity
operator ($\Pi$), the Lorentz generators $J^{i}=\frac{1}{2}%
\epsilon^{ijk}M^{jk}$ and $K^{i}=M^{0i}$ and the chirality operator $\chi$
entering the projectors on the chiral subspaces $(1,0)$ and $(0,1)$ which
satisfy 
\begin{equation}
\{\chi,\Pi\}=0, \qquad [\chi,M^{\mu\nu}]=0, \qquad \chi^2=\mathbf{1}.
\label{chipim}
\end{equation}
The $S^{\mu\nu}$ tensor is the covariant version of the rest-frame parity
operator ($\Pi$) such that $S^{00}=\Pi $ and other components can be written
as 
\begin{equation}
S^{\mu\nu}=\Pi\left( g^{\mu\nu}-i(g^{0\mu}M^{0\nu}+g^{0\nu}M^{0\mu
})-\{M^{0\mu},M^{0\nu}\}\right) .
\end{equation}
This is a symmetric traceless ($S^{\mu}_{~\mu}=0$) tensor with nine
independent components. As a consequence of  Eqs.(\ref{chipim}) we get 
\begin{equation}
\{\chi,S^{\mu\nu}\}=0.   \label{chis}
\end{equation}
The $C$ tensor is given by 
\begin{equation}
C_{\mu\nu\alpha\beta}=4\{M^{\mu\nu},M^{\alpha\beta}\}+2\{M^{\mu\alpha},M^{%
\nu\beta}\}-2\{M^{\mu\beta},M^{\nu\alpha}\}-8(g_{\mu\alpha}g_{\nu\beta
}-g_{\mu\beta}g_{\nu\alpha}).  \label{ctensor}
\end{equation}
with the symmetry properties $C_{\mu\nu\alpha\beta} =-C_{\nu\mu\alpha\beta}
=-C_{\mu\nu\beta\alpha}$ ; $C_{\mu\nu\alpha\beta} =C_{\alpha\beta\mu\nu}$.
It satisfies the Bianchi identity $C_{\mu\alpha\beta\nu}+C_{\mu\beta\nu%
\alpha}+C_{\mu\nu\alpha\beta}=0$ and the contraction of any pair of indices
vanishes $C_{\quad\nu\alpha\beta}^{\nu}=0$. These constraints leave only $10$
independent components. Clearly it satisfies $[\chi,C^{\mu\nu\alpha\beta}]=0$%
.

The covariant basis is orthogonal with respect to the scalar product defined
as $\langle A \vert B\rangle=Tr (AB)$, thus these matrices satisfy the
following relations 
\begin{eqnarray}
Tr\left( \chi\right) =Tr\left( S\right) &=&Tr\left( M\right) =Tr\left( \chi
S\right) =Tr\left( C\right) =0,  \notag \\
Tr\left( \chi M\right) =Tr\left( \chi C\right) =Tr\left( MS\right) =Tr\left(
M\chi S\right) &=&Tr\left( MC\right) =Tr\left( S\chi S\right) =Tr\left(
SC\right) =Tr\left( \chi SC\right) =0.
\end{eqnarray}
where we suppressed the Lorentz indices.

Calculations in this work requires traces of products of the $S^{\mu\nu}$
tensor and other elements in the covariant basis. Let us consider first 
\begin{equation}
Tr\left( SMM\right) =Tr\left( \chi^{2}SMM\right) =-Tr\left( \chi S\chi
MM\right) =-Tr\left( \chi SMM\chi\right) =-Tr\left( SMM\right) \Rightarrow
Tr\left( SMM\right) =0,
\end{equation}
where we used Eqs. (\ref{chipim},\ref{chis}) and the cyclic property of a
trace. Since $\chi$ commutes also with $C$, this procedure can be used to
show that in general if we have a term with an odd numbers of $S$ tensors
the trace of this term will vanish%
\begin{equation}
Tr(\text{term with an odd \# of }S\text{'s})=0.
\end{equation}

The trace of terms with an even number of $S$ factors can always be reduced
to a linear combination of terms with the trace of the product of two $S$ or
two $M$ factors using the following (anti)commutation relations 
\begin{align}
\lbrack M^{\mu\nu},M^{\alpha\beta}] & =-i\left(
g^{\mu\alpha}M^{\nu\beta}-g^{\nu\alpha}M^{\mu\beta}
-g^{\mu\beta}M^{\nu\alpha}+g^{\nu\beta}M^{\mu\alpha}\right)  \label{trcmm} \\
\lbrace M^{\mu\nu},M^{\alpha\beta}\rbrace & = \frac{4}{3}(g^{\mu\alpha}g^{%
\nu\beta}- g^{\mu\beta}g^{\nu\alpha} ) -\frac{4}{3} i
\varepsilon^{\mu\nu\alpha\beta}\chi +\frac{1}{6} C^{\mu\nu\alpha\beta},
\label{tramm} \\
\lbrack M^{\mu\nu},S^{\alpha\beta}\rbrack & =-i\left(
g^{\mu\alpha}S^{\nu\beta
}-g^{\nu\alpha}S^{\mu\beta}+g^{\mu\beta}S^{\nu\alpha}-g^{\nu\beta}S^{\mu%
\alpha}\right) ,  \label{trcms} \\
\left\{ M^{\mu\nu},S^{\alpha\beta}\right\} &
=\varepsilon^{\mu\nu\sigma\beta}\chi S_{\quad\sigma}^{\alpha}
+\varepsilon^{\mu\nu\sigma\alpha}\chi S_{\quad\sigma}^{\beta},  \label{trams}
\\
\lbrack S^{\mu\nu},S^{\alpha\beta}\rbrack & =-i\left(
g^{\mu\alpha}M^{\nu\beta}+g^{\nu\alpha}M^{\mu\beta}
+g^{\nu\beta}M^{\mu\alpha}+g^{\mu\beta}M^{\nu\alpha}\right),  \label{trcss}
\\
\left\{ S^{\mu\nu},S^{\alpha\beta}\right\} & =\frac{4}{3}\left(
g^{\mu\alpha}g^{\nu\beta}+g^{\nu\alpha}g^{\mu\beta} -\frac{1}{2}%
g^{\mu\nu}g^{\alpha\beta}\right) -\frac{1}{6}\left(
C^{\mu\alpha\nu\beta}+C^{\mu\beta\nu\alpha}\right) .  \label{trass}
\end{align}

The simplest case appears in the calculation of $H\to \bar{D}D$ 
\begin{equation}
Tr\left( S^{\mu\nu}S^{\alpha\beta} \right)=Tr\left( \frac{1}{2}%
[S^{\mu\nu},S^{\alpha\beta}]+ \frac{1}{2}\{S^{\mu\nu},S^{\alpha\beta}\}
\right) = 4\left( g^{\mu\alpha}g^{\nu\beta}+g^{\mu\beta}g^{\nu\alpha}-\frac{1%
}{2}g^{\mu\nu}g^{\alpha\beta}\right) \equiv4T^{\mu\nu\alpha\beta}.
\label{trss}
\end{equation}
Similarly, the calculation of $Z^0\to \bar{D}D$ requieres 
\begin{equation}
Tr\left( M^{\mu\nu}M^{\alpha\beta} \right)=Tr\left( \frac{1}{2}%
[M^{\mu\nu},M^{\alpha\beta}]+ \frac{1}{2}\{M^{\mu\nu},M^{\alpha\beta}\}
\right)
=4(g^{\mu\alpha}g^{\nu\beta}-g^{\mu\beta}g^{\nu\alpha})\equiv4G^{\mu\nu%
\alpha\beta}.  \label{trmm}
\end{equation}
The first example of the reduction mentioned above is faced in the
calculation of $Z^0\to \bar{D}D$ which also requires to calculate 
\begin{align}
Tr\left( S^{\mu\nu}S^{\alpha\beta}M^{\rho\sigma}\right) & =Tr\left( \frac{1}{%
2}\left\{ S^{\mu\nu},S^{\alpha\beta}\right\} M^{\rho\sigma} + \frac{1}{2}%
\left[ S^{\mu\nu},S^{\alpha\beta}\right] M^{\rho\sigma} \right)  \notag \\
& =\frac{-i}{2}Tr\left( \left(
g^{\mu\alpha}M^{\nu\beta}+g^{\nu\alpha}M^{\mu\beta}
+g^{\nu\beta}M^{\mu\alpha}+g^{\mu\beta}M^{\nu\alpha}\right) M^{\rho\sigma}
\right)  \notag \\
& =-2i
\left(g^{\mu\alpha}G^{\nu\beta\rho\sigma}+g^{\nu\alpha}G^{\mu\beta\rho%
\sigma}
+g^{\nu\beta}G^{\mu\alpha\rho\sigma}+g^{\mu\beta}G^{\nu\alpha\rho\sigma}
\right).
\end{align}
and 
\begin{align}
Tr\left( S^{\alpha\beta}M^{\mu\nu}S^{\rho\sigma}M^{\gamma\delta}\right) &=
Tr\left( (\frac{1}{2} [S^{\alpha\beta},M^{\mu\nu}] +\frac{1}{2}
\{S^{\alpha\beta},M^{\mu\nu}\}) (\frac{1}{2}[
S^{\rho\sigma},M^{\gamma\delta}] + \frac{1}{2}\{
S^{\rho\sigma},M^{\gamma\delta}\} ) \right)  \notag \\
&=Tr\left( \left( \frac{i}{2}(
g^{\mu\alpha}S^{\nu\beta}-g^{\nu\alpha}S^{\mu\beta}+g^{\mu\beta}S^{\nu%
\alpha}-g^{\nu\beta}S^{\mu\alpha} ) -\varepsilon^{\mu\nu\tau\beta}\chi
S_{\quad\tau}^{\alpha} -\varepsilon^{\mu\nu\tau\alpha}\chi
S_{\quad\tau}^{\beta} \right) \right.  \notag \\
&\left. \left( \frac{i}{2}( g^{\gamma\rho}S^{\delta\sigma} -
g^{\delta\rho}S^{\gamma\sigma}
+g^{\gamma\sigma}S^{\delta\rho}-g^{\delta\sigma}S^{\gamma\rho} )
-\varepsilon^{\gamma\delta\lambda\sigma}\chi S_{\quad\lambda}^{\rho}
-\varepsilon^{\gamma\delta\lambda\rho}\chi S_{\quad\lambda}^{\sigma} \right)
\right)  \notag \\
&= - g^{\mu\alpha} g^{\gamma\rho} T^{\nu\beta\delta\sigma} + g^{\mu\alpha}
g^{\delta\rho} T^{\nu\beta\gamma\sigma} - g^{\mu\alpha} g^{\gamma\sigma}
T^{\nu\beta\delta\rho} + g^{\mu\alpha} g^{\delta\sigma}
T^{\nu\beta\gamma\rho}  \notag \\
&+ g^{\nu\alpha} g^{\gamma\rho} T^{\mu\beta\delta\sigma} - g^{\nu\alpha}
g^{\delta\rho} T^{\mu\beta\gamma\sigma} + g^{\nu\alpha} g^{\gamma\sigma}
T^{\mu\beta\delta\rho} - g^{\nu\alpha} g^{\delta\sigma}
T^{\mu\beta\gamma\rho}  \notag \\
& - g^{\mu\beta} g^{\gamma\rho} T^{\nu\alpha\delta\sigma} + g^{\mu\beta}
g^{\delta\rho} T^{\nu\alpha\gamma\sigma} - g^{\mu\beta} g^{\gamma\sigma}
T^{\nu\alpha\delta\rho} + g^{\mu\beta} g^{\delta\sigma}
T^{\nu\alpha\gamma\rho}  \notag \\
& + g^{\nu\beta} g^{\gamma\rho} T^{\mu\alpha\delta\sigma} - g^{\nu\beta}
g^{\delta\rho} T^{\mu\alpha\gamma\sigma} + g^{\nu\beta} g^{\gamma\sigma}
T^{\mu\alpha\delta\rho} - g^{\nu\beta} g^{\delta\sigma}
T^{\mu\alpha\gamma\rho}  \notag \\
& -4 \left(\varepsilon^{\mu\nu\tau\beta}
\varepsilon^{\gamma\delta\lambda\sigma} T\indices{^{\alpha}_{\tau}^{\rho}_{%
\lambda} } +\varepsilon^{\mu\nu\tau\beta}\varepsilon^{\gamma\delta\lambda%
\rho} T\indices{^{\alpha}_{\tau} ^{\sigma}_{\lambda} }\right.  \notag \\
&
\left.+\varepsilon^{\mu\nu\tau\alpha}\varepsilon^{\gamma\delta\lambda\sigma}
T\indices{^{\beta}_{\tau}^{\rho}_{\lambda} } +\varepsilon^{\mu\nu\tau\alpha}
\varepsilon^{\gamma\delta\lambda\rho} T\indices{^{\beta}_{\tau}
^{\sigma}_{\lambda} }\right)
\end{align}

Similarly it can be shown that 
\begin{align}
Tr\left( M^{\mu\nu}M^{\alpha\beta}M^{\rho\sigma}\right) & =-2i\left(
g^{\mu\alpha}G^{\nu\beta\rho\sigma}-g^{\nu\alpha}G^{\mu
\beta\rho\sigma}-g^{\mu\beta}G^{\nu\alpha\rho\sigma}+g^{\nu\beta}G^{\mu
\alpha\rho\sigma}\right) \\
Tr\left( \chi S^{\gamma\delta}S^{\alpha\beta}M^{\mu\nu}\right) & =-2\left(
\varepsilon^{\mu\nu\sigma\beta}T_{\qquad\sigma}^{\gamma
\delta\alpha}+\varepsilon^{\mu\nu\sigma\alpha}T_{\qquad\sigma}^{\gamma\delta%
\beta}\right) , \\
Tr\left( \chi M^{\mu\nu}M^{\alpha\beta}\right)
&=-4i\varepsilon^{\mu\nu\alpha\beta}.
\end{align}

The calculation of the trace of terms involving six or eight $S$ or $M$
factors (with an even number of $S$ factors) needed in this paper are
reduced in a similar way.

There is a simpler way to obtain these results however, which is specially
useful for terms with six or more factors. Since the result rests only on
the algebraic properties in Eqs. (\ref{trcmm}, \ref{tramm},\ref{trcms},\ref%
{trams},\ref{trcss},\ref{trass}) we can use any representation of these
operators for the calculation of the trace. In this concern the use of the
representation where the internal matrix indices transform as Lorentz
indices is convenient, since in this case the calculation of the trace
reduces to contractions of Lorentz indices which can be easily done using
conventional algebraic manipulation codes like FeynCalc. In this
representation, each internal matrix index $a$ is replaced by a pair of
antisymmetric Lorentz indices $\alpha\beta$ \cite{DelgadoAcosta:2012yc}. The
explicit form of the operators in the covariant basis is given by 
\begin{eqnarray}  \label{basistl}
\left( \mathbf{1}\right)_{\alpha\beta\gamma\delta} &=&\frac{1}{2}
(g_{\alpha\gamma}g_{\beta\delta}-g_{\alpha\delta}g_{\beta\gamma}), \\
\left( \chi \right)_{\alpha\beta\gamma\delta} &=& \frac{i}{2}%
\varepsilon_{\alpha\beta\gamma\delta}, \\
\left( M\indices{_\mu_\nu}\right) _{\alpha\beta\gamma\delta} &=& -i\left(
g_{\mu\gamma}\mathbf{1}_{\alpha\beta\nu\delta}+g_{\mu\delta}\mathbf{1}
_{\alpha\beta\gamma\nu}-g_{\gamma\nu}\mathbf{1}_{\alpha\beta\mu\delta
}-g_{\delta\nu}\mathbf{1}_{\alpha\beta\gamma\mu}\right) , \\
\left( S\indices{_\mu_\nu}\right) _{\alpha\beta\gamma\delta} &=& g_{\mu\nu } 
\mathbf{1}_{\alpha\beta\gamma\delta}-g_{\mu\gamma}\mathbf{1}_{\alpha\beta
\nu\delta}-g_{\mu\delta}\mathbf{1}_{\alpha\beta\gamma\nu}-g_{\gamma\nu }%
\mathbf{1}_{\alpha\beta\mu\delta}-g_{\delta\nu}\mathbf{1}_{\alpha\beta\gamma%
\mu} .
\end{eqnarray}
The explicit form of $C^{\mu\nu\alpha\beta}$ can be constructed from Eq.(\ref%
{ctensor}) and the above relations.

\bibliographystyle{prsty}
\bibliography{dm}

\end{document}